\documentclass[aps,rmp,superscriptaddress,showpacs,showkeys,twocolumn,eqsecnum]{revtex4}

\usepackage{amsmath,amssymb,amsbsy}
\usepackage{graphicx}
\usepackage[usenames]{color}
\usepackage{hyperref}
\usepackage{slashed}
\usepackage{slashed,amsmath}

\newcommand{\cor}[1]{#1}
\newcommand{\del}[1]{}
\newcommand{\new}[1]{{#1}}
\newcommand{\newla}[1]{#1}
\newcommand{\comment}[1]{}
\newcommand{\fig}[1]{fig.~#1}

\newcommand{\M}{\mathcal{M}}

\newcommand{\MeV}{\; \mathrm{MeV} }

\newcommand{\GeV}{\; \mathrm{GeV} }
\newcommand{\ekin}{E_{\mathrm{kin}}}

\newcommand{\BUU}{GiBUU }

\newcommand{\JQE}{J_{\mbox{\begin{tiny}QE\end{tiny}}}}
\newcommand{\SQE}{\sigma^{\mbox{\begin{tiny}QE\end{tiny}}}}

\newcommand{\g}[1]{\gamma^{#1}}
\newcommand{\gF}{\gamma^5}

\newcommand{\tw}[1]{{#1}}
\newcommand{\pr}[1]{{#1}'}

\allowdisplaybreaks
\begin{document}

\title{Exciting pions in nuclei: DCX and Electroproduction in the resonance region} 

\author{O. Buss}
\email{oliver.buss@theo.physik.uni-giessen.de}
\homepage{http://theorie.physik.uni-giessen.de/~oliver}
\thanks{Work supported by DFG.}

\affiliation{Institut f\"ur Theoretische Physik, Universit\"at Giessen, Germany}

\author{L. Alvarez-Ruso}
\affiliation{Departamento de F\'{\i}sica Te\'orica and IFIC,
Universidad de Valencia - CSIC, Spain}

\author{A.~B. Larionov}

\altaffiliation{Also in RRC "I.V. Kurchatov Institute",
                       123182 Moscow, Russia. \\Present address: Frankfurt Institute for Advanced Studies (FIAS), Germany.}

\author{U. Mosel}

\affiliation{Institut f\"ur Theoretische Physik, Universit\"at Giessen, Germany}


\date{\today}
\begin{abstract}
We describe the double charge exchange (DCX) reaction of pions with different nuclear targets at incident kinetic energies of $120-180$~MeV within the GiBUU transport framework. The DCX process  is highly sensitive to details of the interactions of pions with the nuclear medium and, therefore, represents a major benchmark for any model of pion scattering off nuclei at low and intermediate energies. We achieve a good quantitative agreement with the extensive data set obtained at LAMPF. Furthermore, we present first results on electron scattering off nuclei  for beam energies of 700-1500 MeV and virtualities of $Q^2\leq 700 \MeV^2$. Including quasi-elastic and pion-production processes and employing in-medium kinematics and realistic form factors, we evaluate inclusive cross sections and compare to data obtained at the ADONE storage ring~(Frascati).
\end{abstract}

\maketitle
\section{Introduction and Motivation}\label{intro}
Major experimental efforts are being undertaken to understand the in-medium properties of hadrons, e.g.~\cite{Messch,Trnka:2005ey, Adamova:2006nu}. \del{One can distinguish the performed experiments in two categories: the ones using elementary targets scattering off a nucleus which remains close to its ground state, e.g.~\cite{Messch,Trnka:2005ey}, and the ones who employ heavy ion beams with the consequence of a strongly time-dependent medium, e.g.~\cite{Adamova:2006nu}. }In many cases, the hadrons of interest decay inside the nuclear medium while one finally observes its reaction products outside the medium. However, one must discriminate between the change of the original hadron properties in the medium and many-body final state re-scattering effects which influence the observed decay products in the case of the hadronic observables. In this context, transport models become relevant: these models describe the whole reaction process in a microscopical manner by evolving the phase space density of all involved particle species in time. Therefore, the transport approach is suited to shed light on this discrimination problem.

The Giessen Boltzmann-Uehling-Uhlenbeck (GiBUU) transport model~\cite{GiBUUWebpage} has been developed over the last 20 years to describe heavy ion collisions, photon-, electron-, pion- and neutrino-induced reactions within one unified transport framework. After a major restructuring phase of the whole source code, we report here new applications of this method.

In this work we focus on low-energy processes, where pions play a dominant role. Recently, we have shown using GiBUU~\cite{Muhlich:2004zj,Buss:2006vh} that pion re-scattering in the final state description of photon induced double-pion production produces a considerable modification of the $\pi\pi$  invariant mass distributions observed by the TAPS collaboration~\cite{Messch}. The propagation of low-energy pions in nuclear matter in the GiBUU framework has already been extensively discussed in ref.~\cite{Buss:2006vh} and compared to quantum mechanical calculations. In the spirit of benchmarking the model, pionic double charge exchange (DCX) is another very interesting reaction. The fact that DCX requires at least two nucleons to take place makes it a very sensitive test for pion re-scattering and absorption. 

We also present our first results for electron scattering off nuclei for beam energies between $700\MeV$ and $1500\MeV$ and virtualities of $Q^2\leq 632 \MeV^2$. Employing impulse approximation, we obtain inclusive cross sections with full in-medium kinematics and realistic form factors for quasi-elastic scattering and pion production.

This article is structured in the following way. First we introduce our GiBUU transport model emphasizing the most relevant issues. Next, we discuss two distinct applications: DCX and electron scattering off nuclei.

\section{The \BUU transport model}\label{buu}
Boltzmann-Uehling-Uhlenbeck (BUU) transport models are based on the Boltzmann equation, which was modified  by Nordheim, Uehling and Uhlenbeck to incorporate quantum statistics. A brief description of the formalism is given below. For a detailed discussion concerning the physical input for pion-induced reactions we refer the reader to~\cite{Buss:2006vh,Buss:2006yk} and the references therein. 

\subsection{The BUU equation}\label{buuEQ}
The BUU equation actually consists of a series of coupled differential equations, which describe the time evolution of the \newla{single}-particle \newla{phase-space densities $f_a(\vec{r},\vec{p},t)$. The index $a=\pi,\omega,N,\Delta,\ldots$ denotes the different particle species in our model.} A large number of mesonic and baryonic states is actually included, but at the energies of interest for this study, the relevant ones are $\pi,N$ and the $\Delta(1232)$ resonance.

For a particle of species X, its time evolution is given by
\begin{eqnarray}
&&\frac{d f_X(\vec{r},\vec{p},t)}{d t}=\frac{\partial f_X(\vec{r},\vec{p},t)}{\partial t} +\frac{\partial H_X}{\partial \vec{p}}\frac{\partial f_X(\vec{r},\vec{p},t)}{\partial \vec{r}} \nonumber
\\  &&-\frac{\partial H_X}{\partial \vec{r}}\frac{\partial f_{X}(\vec{r},\vec{p},t)}{\partial \vec{p}}   
=I_{coll}\left(f_X,f_a,f_b,\ldots\right)
\label{BUUEquation}
\end{eqnarray}
with the one-body Hamilton function 
\begin{eqnarray}
H_X(\vec{r},\vec{p})&=&\sqrt{\left(\vec{p}-\vec{A}_X(\vec{r},\vec{p})\right)^{2}+m_X^{2}+U_X(\vec{r},\vec{p})}\nonumber \\
&& +A_X^{0}(\vec{r},\vec{p}) \; .
\label{HamiltonFunc}
\end{eqnarray}
The scalar potential $U_X$ and the vector potential $A^\mu_X$ of species X may in principle depend upon the phase space densities of all other species. Hence, the differential equations are already coupled through the mean fields. In the limit of $I_{coll}=0$, eq.~(\ref{BUUEquation}) becomes the well-known Vlasov equation. The collision term $I_{coll}$ on the right-hand side of eq.~(\ref{BUUEquation}) incorporates explicitly all scattering processes among the particles. The  reaction probabilities used in this collision term are chosen to match the elementary collisions among the particles in vacuum. Therefore, we model these probabilities as a sum of dominant resonance terms and small background terms. The latter resonance terms are based on the partial wave analysis of Manley et al.~\cite{ManleySaleski}. The background terms are defined as differences between experimental data and resonance contributions. Note that background contributions are instantaneous in space-time, whereas the resonances propagate along their classical trajectories until they decay or interact with one or two nucleons in the medium. 

Concerning pion absorption, the most important mechanisms are a two step process in which $\pi N \to \Delta$ is followed by $\Delta N \to NN$ \new{or $\Delta N N \to NNN$}, and the one step background process $\pi N N \to N N$. 

\subsection{Medium effects}
\label{mediumEffects}
As medium modifications, we include Pauli-blocking and Fermi-motion of the nucleons and Coulomb forces. Furthermore, we account for the collisional broadening in the $\Delta$ resonance width. The hadronic potentials included are introduced as \cor{time-like} components of vector potentials in the local rest-frame~\cite{Teis:1996kx}. For our purposes the most important mean field potentials are those acting on the nucleon and the $\Delta$ resonance. The nucleon mean-field potential is parametrized by Welke et al.~\cite{Welke} as a sum of a Skyrme term depending only on density and a momentum-dependent contribution (for explicit details and parameters see~\cite{Teis:1996kx}). Phenomenology tells us that the $\Delta$ potential  has a depth of about $-30 \MeV$ at $\rho_{0}$\,\cite{ericsonWeise,Peters:1998mb}. Comparing to a momentum independent nucleon potential, which is approximately $-50 \MeV$ \cor{deep}, the $\Delta$ potential is, \cor{therefore,} taken to be
\begin{eqnarray*}
        A^0_{\Delta}(\vec{p},\vec{r})=\frac{2}{3}\ A^0_{\mathrm{nucleon}}(\vec{p},\vec{r}).
        \label{DeltaPotential}
\end{eqnarray*}
Here we \cor{assume} the same momentum dependence for the nucleon and the $\Delta$ potential. No potential for the pion was included in the calculations.

\section{Pionic double-charge-exchange in nuclei}\label{resu}
\subsection{Introduction}
The double charge exchange (DCX) reaction of pions received a considerable attention in the past (see for instance Ref.~\cite{LAMPF} and references therein). The mechanism of two sequential single charge exchanges has traditionally been able to explain the main features of this reaction~\cite{Becker:1970tk,Gibbs:1977yz} at low energies although the contribution of the $A(\pi,\pi\pi)X$ reaction becomes progressively important as the energy increases~\cite{Vicente:1988iv,Alqadi:2001pe}. At higher ($\sim1~\mathrm{GeV}$) energies, the sequential mechanism becomes insufficient to account for the reaction cross section~\cite{Abramov:2002nz,Krutenkova:2005nh}.
Extensive experimental studies performed at LAMPF obtained high precision data for doubly differential cross sections on $^3He$~\cite{Yuly:1997ja} and heavier nuclei ($^{16}O$, $^{40}Ca$, $^{208}Pb$)~\cite{Wood:1992bi} in the region of $\ekin=120-270 \MeV$.  

H\"ufner and Thies~\cite{huefnerThies} explored for the first time the applicability of the Boltzmann equation in $\pi A$ collisions and achieved qualitative agreement with data on single and double charge exchange. Their method to solve the Boltzmann equation was based upon an expansion of the pion one-body distribution function in the number of collisions. There, in contrast to our work, the Boltzmann equation is not solved with a test-particle ansatz but by reformulating it into a set of coupled differential equations which can then be solved in an iterative manner. However, this approach was based on simplifying assumptions of averaged cross sections and averaged potentials.
The work by Vicente et al.~\cite{Vicente:1988iv} was based upon the cascade model described in~\cite{osetSimulation}. There, a microscopic model for $\pi N$ scattering was used as input for the pion reaction rates in the simulation. In that work~\cite{Vicente:1988iv}, pion DCX off $^{16}O$ and $^{40}Ca$ has been explored and fair quantitative agreement with data has been achieved.

In the following, we explore DCX on heavier nuclei, comparing with the data measured by Wood et al.~\cite{Wood:1992bi}. We also address the scaling of the total cross section discussed by Gram et al.~\cite{Gram:1989qh}. To focus only on single-pion re-scattering, we consider incoming pion energies below $\ekin=180\MeV$; above that energy $2\pi$ production becomes prominent and DCX does not happen necessarily in a two-step process anymore. 
Due to the small mean free path of the incoming pions, the process is mostly sensitive to the surface of the nucleus. In \cite{Buss:2006yk} we also addressed the issue of parallel versus full ensemble scheme in the numerical implementation of the BUU equation. There, we found that the parallel ensemble scheme is a good approximation and has therefore been applied in the following calculations.

\subsection{Influence of the density profile} \label{densChapter}
\begin{figure}[b]
\begin{center}
\includegraphics[]{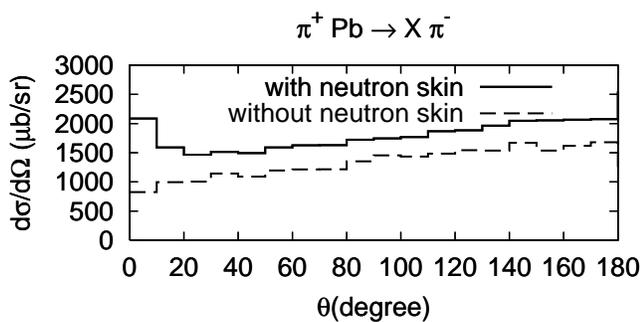}
\label{densVer}
\caption{Influence of the density distribution on the angular distributions for the double charge exchange process $\pi^+ Pb\to \pi^- X$ at $E_{\mathrm{kin}}=180 \MeV$. The solid line shows the result where we implement different radii of the density distributions of neutrons and protons in our model ($R_n-R_p=0.266$ fm). As a consequence, the nucleus is surrounded by a small neutron skin. For the dashed line we assumed same radii for the proton and neutron distributions.}
\end{center}
\end{figure}
Due to the low pion mean free path in nuclear matter~\cite{Buss:2006vh}, DCX is very sensitive to the surface properties of the nuclei. Therefore, we compared the results with a neutron skin for the $Pb$ nucleus to the results obtained without such a neutron skin; for details see \cite{Buss:2006yk}. 
Neutron skins are very interesting because in those skins only $\pi^+$ mesons can undergo charge exchange reactions. For the positive pions this causes an enhancement of DCX processes at the surface, so the pions do not need to penetrate deeply for this reaction. Hence, the probability for their absorption is reduced. As can be observed in fig. 1, the enhancement in the total cross section for $R_n-R_p=0.266$ fm is roughly $35\%$ at $180\MeV$. 
The accurate determination of neutron skins is relevant for different 
areas of physics such as nuclear structure, neutron star properties, 
atomic parity violation (PV) and heavy ion collisions \cite{Horowitz:1999fk,Horowitz:2006iv,Piekarewicz:2006vp}. The 
Parity Radius Experiment (PREX) at JLab \cite{PREX} shall measure the neutron 
radius with high precision ($1\%$) using PV electron scattering. We can see that the DCX cross section is very sensitive to the size of the 
difference between the proton and the neutron radii of $Pb$. The effect is 
specially large (more than a factor $2$) at forward angles, where our model 
performs very well (see next Section). Indeed, without neutron skin, due to strong pion absorption in the bulk of nuclear matter, the DCX cross section is small at forward angles. The presence of a neutron skin favors DCX in peripheral reactions, where the pion propagates in practically pure neutron matter. This naturally enhances the DCX cross section at forward angles.
Hence, a precise measurement of DCX at forward angles, combined with a realistic theoretical analysis could be a valuable source of information on the neutron skins complementary to the one obtained with PV electron scattering. Note that for the $\pi^-$ similar arguments lead to a reduction of the cross section.



\subsection{Comparison to data}
\begin{figure}[]
\begin{center}
\includegraphics[width=0.45\textwidth]{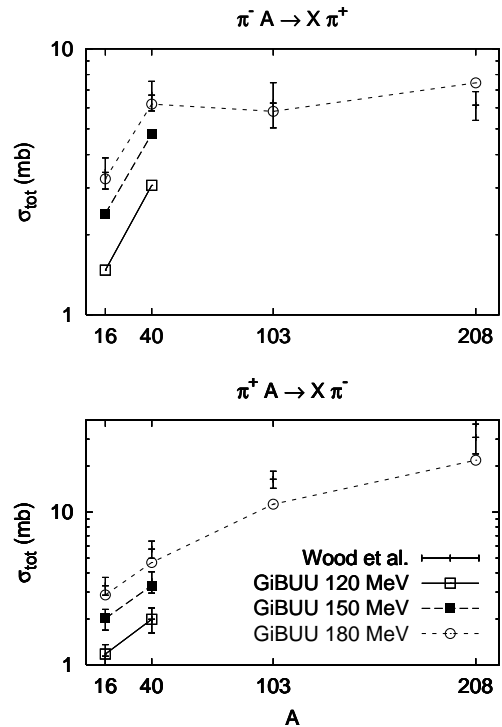}
\end{center}
\caption{\new{The inclusive double charge exchange total cross section as function of the nuclear target mass at $E_{\mathrm{kin}}=120,150$ and $180 \MeV$. The lines connecting our results are meant to guide the eye; the data are taken from Ref.~\cite{Wood:1992bi} (lower panel: $E_{\mathrm{kin}}=120,150$ and $180  \MeV$, upper panel: only $180 \MeV$).}}
\label{total}
\end{figure}
Now we proceed to the comparison with the data measured at LAMPF by Wood et al.~\cite{Wood:1992bi}. We discuss first the total cross section and thereafter the angular distributions. \footnote{In \cite{Buss:2006yk} we also show double differential cross sections as a function of both angles and energies of the outgoing pions.}

In \fig{\ref{total}} one can see the excellent quantitative agreement to the total cross \newla{section} data at $120$, $150$ and $180 \MeV$ for Oxygen and Calcium. For the Lead nucleus we see some discrepancies, which are, however, within the experimental accuracy. Notice that we reproduce the different $A$ \new{dependencies} of \new{both} $(\pi^+,\pi^-)$ and $(\pi^-,\pi^+)$ reactions. It is due to the fact that, when $A$ increases, the number of neutrons increases with respect to the number of protons, and this favors the $\pi^+$ induced reaction.

\begin{figure}[]
\begin{center}
\includegraphics[width=0.45\textwidth]{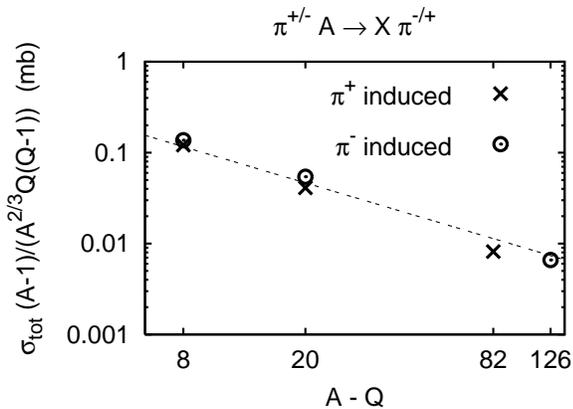}
\end{center}
\caption{The scaling of the total charge exchange cross section according to eq.~(\ref{scalingEQ}) is visualized by dividing $\sigma_{tot}$ by the factor $A^{2/3}Q(Q-1)/(A-1)$  and plotting it as a function of $A-Q$. $Q$ denotes the number of protons in the case of $(\pi^-,\pi^+)$ and the number of neutrons in $(\pi^+,\pi^-)$. The points are GiBUU results at pion kinetic energies of $180 \MeV$; the dashed line
denotes a function proportional $1/(A-Q)$, corresponding to the exact scaling.}
\label{scaling}
\end{figure}
\begin{figure}[h]
\begin{center}
\includegraphics[width=0.4\textwidth]{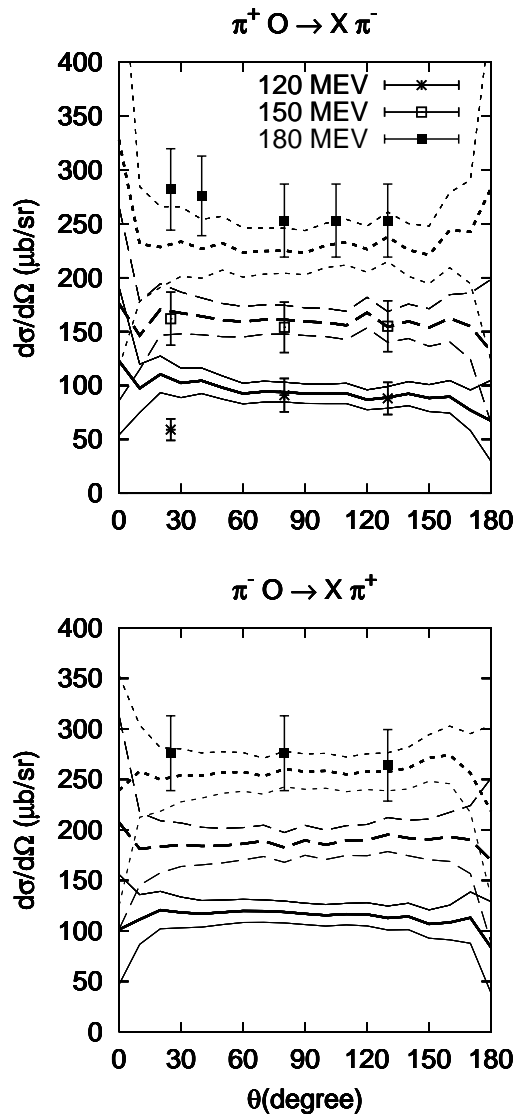}
\end{center}
\caption{Angular distributions for the double charge exchange process $\pi^\pm A\to \pi^\mp X$ at $E_{\mathrm{kin}}=120,150$ and $180 \MeV$. The data points are taken from \cite{Wood:1992bi}; only systematical errors are shown. The solid lines represents the GiBUU results at $120 \MeV$, the dashed lines at $150 \MeV$ and the dotted lines at $180 \MeV$. The bold lines represent the results while the thin lines represent a $1\sigma$ confidence level for each point based upon our statistics.}

\label{parallel}
\end{figure}

\begin{figure}[h]
\begin{center}
\includegraphics[width=0.4\textwidth]{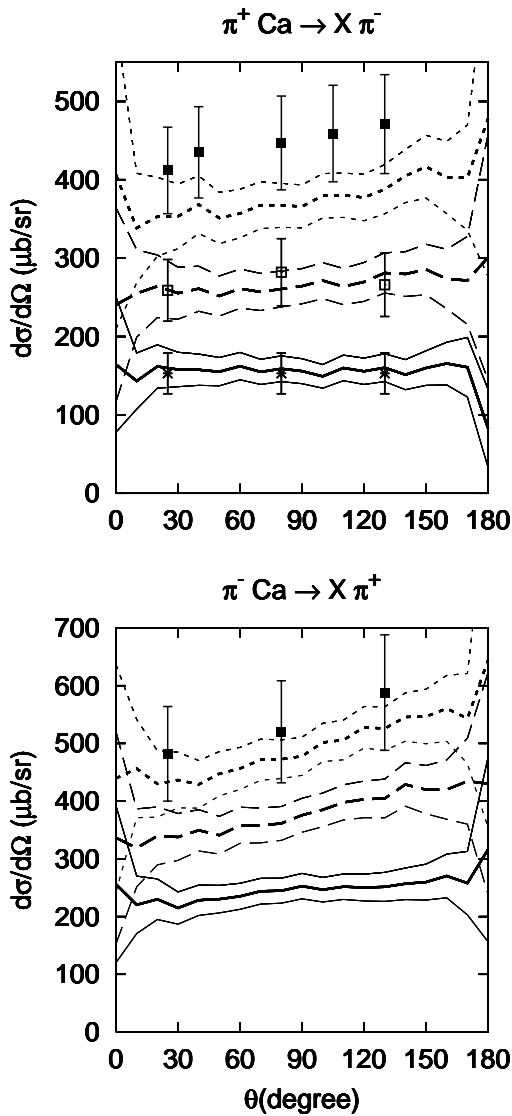}
\end{center}
\caption{Same as fig.\ref{parallel} for $^{40}Ca$.}
\label{parallel1}
\end{figure}

\begin{figure}[h]
\begin{center}
\includegraphics[width=0.45\textwidth]{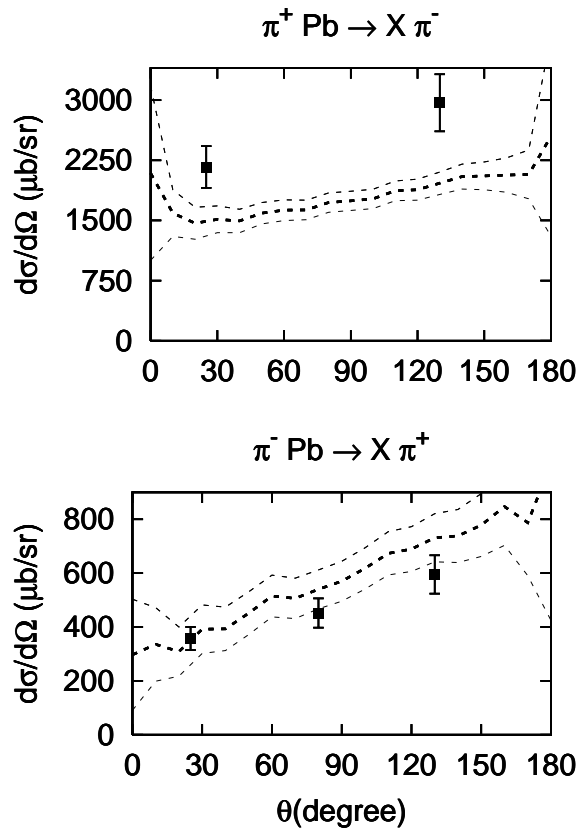}
\end{center}
\caption{Same as fig.\ref{parallel} for $^{208}Pb$.}
\label{parallel2}
\end{figure}
In \cite{Gram:1989qh}, Gram et al. discuss a scaling law of the total cross section. They argue as follows. Since the first collision takes place predominantly at the surface, the cross section should scale with $A^{2/3}$. Furthermore they assume that  DCX is mainly a two-step process, and that a pion which undergoes an elastic process at the first collision will not contribute. This is reasonable because the incoming pions loose energy in the elastic process, and their cross section for a second charge-exchange reaction is hereafter very much reduced. For a negative pion the first charge exchange reaction occurs with a probability of $Z/N$ where $Z(N)$ denotes the number of protons(neutrons). This is the case if the interaction is dominated by the $\Delta$ resonance, as it should be in this energy region. Finally, the second charge exchange process then takes place with the probability $(Z-1)/(A-1)$ since, in the isospin limit, the $\pi^0$ interacts equally well with protons and neutrons.  
Putting these considerations together and extending them to the $\pi^+$ case, the cross section for DCX is expected to scale according to
\begin{eqnarray}
\sigma_{tot}\sim A^{2/3}  \frac{Q}{A-Q}\frac{Q-1}{A-1} \; ,
\label{scalingEQ}
\end{eqnarray}
where $Q$ denotes the number of protons in the case of $\pi^-$ induced and the number of neutrons in $\pi^+$ induced DCX. 

Gram et al. \cite{Gram:1989qh} find good agreement of this scaling law with experimental data. This scaling is fulfilled within our GiBUU simulations as can be seen in \fig{\ref{scaling}}. Nevertheless, one may wonder why this scaling law works in a process which is so sensitive to the neutron skin on heavy nuclei, as has been shown in fig. 1. Since the first collision takes place on the surface, a neutron skin causes an enhancement in the $A(\pi^+,\pi^-)X$ reaction while $A(\pi^-,\pi^+)X$ is suppressed. This effect leads to a deviation from the scaling. However there are also Coulomb forces which are not negligible. The Coulomb force enhances $A(\pi^-,\pi^+)X$ by attracting the negative projectiles and repelling the positive products, which therefore have a smaller path in the nucleus and undergo less absorption. And, due to similar arguments, the reaction $A(\pi^+,\pi^-)X$ is suppressed. We find that this effect counteracts the one from the neutron skin restoring the scaling.  In any case\new{,} the approximate scaling exhibited by the cross section shows that the reaction is very much surface driven and can be very well understood in terms of a two-step process.

In fig. \ref{parallel}-\ref{parallel2} we show $d\sigma / d\Omega$ for DCX at $E_{\mathrm{kin}}=120,150$ and $180 \MeV$ on $^{16}O$, $^{40}Ca$ and $^{208}Pb$ as a function of the scattering  angle $\theta$ in the laboratory frame. Our results 
(bold lines) are shown together with their uncertainties of statistical nature (thin lines). The latter ones are well under control except at very small and very large angles, where statistics is very scarce. Again, there is a very good quantitative agreement for both $O$ and $Ca$. In the $Pb$ case, the $(\pi^-,\pi^+)$ reaction is well described, but the $(\pi^+,\pi^-)$ one is underestimated in spite of the enhancement caused by the neutron skin.

\section{Electron induced processes}\label{electron}
Our investigation of electron induced processes in the resonance regime has two major goals. On the one hand side, we want to investigate the influence of final state events on possible electron signals; and on the other hand side we want to understand the role of the hadronic resonances in the scattering of electrons and nuclei. 

Furthermore, Leitner et al.~\cite{Leitner:2006sp,Leitner:2006ww} have already performed calculations for neutrino-induced reactions in the resonance energy regime using the GiBUU transport model. The electron and neutrino interaction with nuclei is closely interconnected, so we can test assumptions going into the neutrino calculations. This is interesting, since unlike in the electron sector there are almost no experimental data available in the neutrino sector.

The wavelength of the photon is considered small as compared to intra-nucleon distances in the nucleus. Hence we consider the nucleus as a Fermi-Gas of nucleons, and the total reaction rate is given by an incoherent sum over all nucleons. These nucleons are under the influence of a mean field potential as described in \ref{mediumEffects}, therefore the dispersion relation is modified in the medium. However, we neglect at the present stage modifications of the nucleon width due to short range correlations.

\subsection{(Quasi-)Elastic Scattering off a single nucleon}
In this section we will treat the most trivial case of electron-nucleon scattering: the elastic scattering. In the case of electron scattering off a bound nucleon this process is referred as \textit{quasi-elastic (QE)}.

\begin{figure}[]
\begin{center}
 \includegraphics[width=0.2\textwidth]{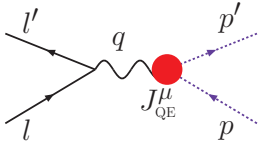}
\end{center}
 \caption{(Quasi-)Elastic electron-nucleon scattering.}
\label{feynDiag_QE}
\end{figure}

Our notation is shown in fig.~\ref{feynDiag_QE}: the in- and outgoing lepton momenta are denoted  $\tw{l}$ and $\pr{l}$, the nucleon momenta $\tw{p}$ and $\pr{p}$, the photon momentum by $q= \tw{l}-\pr{l}$. The $\gamma$ matrices and the spinor normalization are taken according to Bjorken and Drell \cite{BjorkenDrell}.  The electron mass is denoted by $m_e$, the masses of the in- and outgoing nucleons by $\tw{m}$ and $\pr{m}$.

Parametrising the hadronic vertex by the operator $J_{\mbox{\begin{tiny}QE\end{tiny}}}^\mu$, the matrix element for this process, summed over final and averaged over initial spin, is given by
\begin{eqnarray}
\label{QE_ME}
&&|\M|^2
=
\underbrace{\frac{e^2}{2 q^4} \sum_{t_i,t_f}
\overline{v}(\pr{l},t_f) \gamma_{\mu} v(\tw{l},t_i)  
\left( \overline{v}(\pr{l},t_f) \gamma_{\nu} v(\tw{l},t_i) \right)^\dagger}_{=L_{\mu\nu}} \nonumber\\
&&\times 
\underbrace{\frac{1}{2}\sum_{s_i,s_f}
\overline{u}(\pr{p},s_f) (J_{\mbox{\begin{tiny}QE\end{tiny}}})^{\nu} u(\tw{p},s_i) \left(\overline{u}(\pr{p},s_f) (J_{\mbox{\begin{tiny}QE\end{tiny}}})^{\mu} u(\tw{p},s_i)\right)^\dagger}_{=H_{\mbox{\begin{tiny}QE\end{tiny}}}^{\nu\mu}} \nonumber \\
&&=L_{\mu\nu} H_{\mbox{\begin{tiny}QE\end{tiny}}}^{\nu\mu} \; .\nonumber
\end{eqnarray}
with $s_i$, $s_f$ denoting the initial and final nucleon spin; $t_i$, $t_f$ denoting the electron spins. In the latter equation, the electron and nucleon contributions are separated in the so-called hadronic and leptonic tensors $H_{\mbox{\begin{tiny}QE\end{tiny}}}^{\nu\mu}$ and $L_{\mu\nu}$. 

\begin{figure}[]
\begin{center}
 \includegraphics[width=0.2\textwidth]{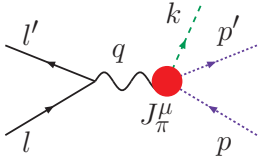}
\end{center}
 \caption{Electron induced pion production off a nucleon.}
\label{feynDiag}
\end{figure}

The most general form of the hadronic vertex can be expressed as
\begin{eqnarray*}
\JQE^\mu=e^2 \left(F_1 \gamma^\mu + \frac{i}{2m_0}F_2 \sigma^{\mu\nu}q_\nu+F_3 q^\mu \; \right)
\end{eqnarray*}
with the nucleon vacuum mass  $m_0\simeq 0.938$ GeV. Demanding charge conservation at the vertex, i.e. $q_\mu j^\mu = 0$, we get
\begin{eqnarray}
\JQE^\mu=e^2\left(F_1 \gamma^\mu + \frac{i}{2m_0}F_2 \sigma^{\mu\nu}q_\nu+\frac{F_1(\tw{m}-\pr{m})}{q^2} q^\mu \right)\; 
\label{generalQE}
\end{eqnarray}
where the last summand is absent in the case of non-momentum-dependent nucleon masses. In the vacuum, the masses of the in- and outgoing nucleon are identical, therefore this latter term vanishes. In the medium, however, the masses depend through the mean fields on the momenta of the nucleons. The form factors $F_1$ and $F_2$ are the standard \textit{Dirac} and \textit{Pauli form factors}. As an approximation, we choose the in-medium form-factors to be the same as the vacuum form factors.

Finally, the form factors  $F_1$ and $F_2$ can be expressed by the so called Sachs form-factors $G_e$ and $G_m$, which have in the electron-nucleon CM-frame an interpretation as charge form factor and magnetic form factor. For the Sachs form-factors $G_e^p$, $G_m^p$,$G_m^n$ we will use the so called \textit{BBA03 parametrization}~\cite{Budd:2003wb}. For $G_e^n$ we use a parametrization obtained by Krutov and Troitsky~\cite{Krutov:2002tp}.
Finally, the cross section for quasi elastic scattering is given by the general formula
\begin{eqnarray}
d\SQE&=&
 \frac{1}{\sqrt{(\tw{l}_\alpha \tw{p}^\alpha)^2}} 
 \frac{m_e^2~ \tw{m}~ \pr{m}}{2\pi^2}d\Omega_{\pr{l}}~\\ && d|\vec{\pr{l}}|~|\vec{\pr{l}}| ~ \delta((l+p-\pr{l})^2-\pr{m}^2)
 ~|\M|^2\nonumber \; .
\end{eqnarray}
Note that we use the full in-medium kinematics, therefore $m\neq m'$ in the general case.

\subsection{Electron induced single-pion production off a single nucleon}
As depicted in fig.~\ref{feynDiag}, we now consider an additional pion with momentum $k$ in the final state.
As in the QE-scattering, the matrix element can be written in the form
\begin{eqnarray}
|\M|^2
&=&L_{\mu\nu} H_{\pi}^{\nu\mu} \;
\end{eqnarray}
where $H_\pi$ is now including another hadronic vertex contribution as compared to the case of QE-scattering:
\begin{eqnarray*}
H_{\pi}^{\nu\mu}
&=&
\frac{1}{2}\sum_{s_i,s_f}
\overline{u}(\pr{p},s_f) (J_\pi)^{\nu} u(\tw{p},s_i) \\ &&\times \left(\overline{u}(\pr{p},s_f) (J_\pi)^{\mu} u(\tw{p},s_i)\right)^\dagger \; .
\end{eqnarray*}
In the same spirit as Berends et al.~\cite{Berends:1967vi} but in the notation of MAID \cite{Pasquini}, we parametrize this hadronic vertex by
\begin{eqnarray}
J_\pi^{\mu}=\sum_{i=1}^6 A_i M_i^\mu \label{flux}
\end{eqnarray}
with
\begin{eqnarray}
M_1^\mu &=&\frac{-i}{2}\gF (\g{\mu}\slashed{q} -\slashed{q} \g{\mu}) \label{Ms} \\
M_2^\mu &=&2 i  \gF (P^\mu q\cdot(k-\frac{q}{2})-P \cdot q (k-\frac{q}{2})^\mu)  \nonumber \\
M_3^\mu &=&  -i \, \gF (\g{\mu} k\cdot q-\slashed{q}k^{\mu}) \nonumber   \\
M_4^\mu &=&-2i \, \gF (\g{\mu} q\cdot P-\slashed{q} P^\mu) -2\tw{m} M_1^\mu  \nonumber\\
M_5^\mu &=&  i  \, \gF (q^\mu k\cdot q - q^2 k^\mu)   \nonumber   \\
M_6^\mu &=&-i   \, \gF (\slashed{q} q^{\mu}-q^2 \g{\mu}) \nonumber 
\end{eqnarray}
and
$
P=(\pr{p}+\tw{p})/2\; .
$
The form factors $A_1,\dots,A_6$, the so called \textit{invariant amplitudes},  are functions depending on all possible scalars which one can construct in the vacuum out of the available $4$-vectors at the vertex. As a minimal set of scalars we choose
$W=\sqrt{s}$, $Q^2=-q_\mu q^\mu$ and the CM scattering angle $\theta$ between $\vec{q}$ and $\vec{k}$.
The $A_i$ are parametrized by the MAID group \cite{MAIDWebsite,Tiator:2006dq,Drechsel:1992pn} according to elementary scattering processes. As a first approximation we will use the same parametrizations also in the medium case. However, we have to be careful since $W$ can get modified in the medium. So we define 
\begin{eqnarray*}
\tw{p}\, ^{\textrm{vacuum}}&=&(\sqrt{m_0^2+ \vec{\tw{p}}^2},\vec{\tw{p}})
\end{eqnarray*}
with $m_0=0.938 \GeV$ and \[
W^{\textrm{vacuum}}= \sqrt{(q+\tw{p}\, ^{\textrm{vacuum}})^\mu(q+\tw{p}\, ^{\textrm{vacuum}})_\mu}\; . \]
We choose the CM-frame of the hadronic vertex as our reference frame for its evaluation. Finally, we define the in-medium form-factors as the vacuum form-factors of the MAID analysis evaluated at the kinematical point $Q^2, W^{\textrm{vacuum}}, \theta$. The cross section in the medium is given by 
\begin{widetext}
\begin{eqnarray*}
&& \frac{d\sigma^\pi_{j}}{d{\pr{l}}^0 d\Omega_{\pr{l}} d\Omega_k} 
=
%
 \left(\frac{1}{|v_e-v_n|} \frac{m_e^2~\pr{m}~\tw{m}}{2~(2\pi)^5}   \dfrac{|\vec{k}|^2~\pr{l}^0}{k^0~ \pr{p}^0~\tw{l}^0~\tw{p}^0}
\frac{L_{\mu\nu}H_{\pi}^{\mu\nu} }{D}
\right)_{|\vec{k}|=x_0} 
\end{eqnarray*}
with
\begin{eqnarray*}
D=\left|\frac{|\vec{k}|}{k^0}+
\frac{
\left(|\vec{k}|-(\vec{q}+\vec{\tw{p}}_j)\vec{k}/|\vec{k}|\right)
\left(1+\frac{1}{|\vec{\pr{p}}|}\left(2 \frac{|\vec{\pr{p}}|}{\sqrt{\vec{\pr{p}}^2+m_0^2}}\; V(|\vec{\pr{p}}|)
+2\sqrt{\vec{\pr{p}}^2+m_0^2}  \frac{d V(|\vec{\pr{p}}|)}{d|\vec{\pr{p}}|} 
+V(|\vec{\pr{p}}|)\frac{dV(|\vec{\pr{p}}|)}{d|\vec{\pr{p}}|}  \right)\right)}{\pr{p}^0} 
\right|
\end{eqnarray*}
\end{widetext}
and $x_0$ is the solution for $|\vec{k}|$ to the energy conservation condition
\begin{eqnarray*}
q^0+\tw{p}^0&=&k^0+\pr{p}^0\\
&=&\sqrt{|\vec{k}|^2+m_\pi^2}+ \sqrt{|\vec{\tw{p}}+\vec{q}-\vec{k}|^2+\left[\pr{m}(\vec{\tw{p}}+\vec{q}-\vec{k})\right]^2} \; .
\end{eqnarray*}
Note that $\tw{m}$ and $\pr{m}$ are the in-medium masses and $V$ denotes the potential of the nucleon, i.e. 
\[
\pr{p}=\left(\sqrt{\vec{\pr{p}}^2+m_0^2}+V(\vec{\pr{p}}),\vec{\pr{p}}\right) \; .
\]

\subsection{Scattering off nuclei}
We assume that the electron scattering off a nucleus takes place in a two-step process. First the electron excites a pion and a nucleon inside the nucleus, this is the so called \textit{initial state}. In the so called \textit{final state} process the particles are transported out of the nucleus.  This may, owing to re-scattering effects, lead to a change in the final state particle multiplicities and distributions. However, we will restrict ourselves in this work to inclusive cross sections and discuss the FSI effects in a forthcoming publication.

Fixing the electron kinematics, the inclusive cross section for scattering off a nucleus is given by
\begin{eqnarray*}
&&\frac{d\sigma_{tot}}{d{\pr{l}}^0 d\Omega_{\pr{l}}}=
\left(
\sum^A_{j=1}\frac{d\SQE_j}{d{\pr{l}}^0 d\Omega_{\pr{l}}} P_{PB}(\vec{r}_j,\vec{\pr{p}})
\right)_{\pr{p}=l+p-l^\prime}
\\
&&+\left( \int  d\Omega_k 
 \sum^A_{j=1}\frac{d\sigma^\pi_{j}}{d{\pr{l}}^0 d\Omega_{\pr{l}} d\Omega_k}  P_{PB}(\vec{r}_j,\vec{\pr{p}}) 
\right)_{\pr{p}=l+p-l^\prime-k}
\end{eqnarray*}
where we sum over all nucleons in the nucleus. Note also, that we have to integrate in the latter equation over the angular distribution of the pion in the $\gamma^\star N\to \pi N$ contribution. The functions $P_{PB}$ denotes the Pauli blocking factor at initialization time. 
Up to now, we do not consider any other contributions than quasi-elastic scattering and single $\pi$ production, i.e. $\pi\pi$ production is omitted. Henceforth, we expect our model to fail above the $\pi\pi$ threshold.
\subsection{Results}
\begin{figure}[h!]
\centering
\includegraphics[width=0.3\textwidth,angle=270]{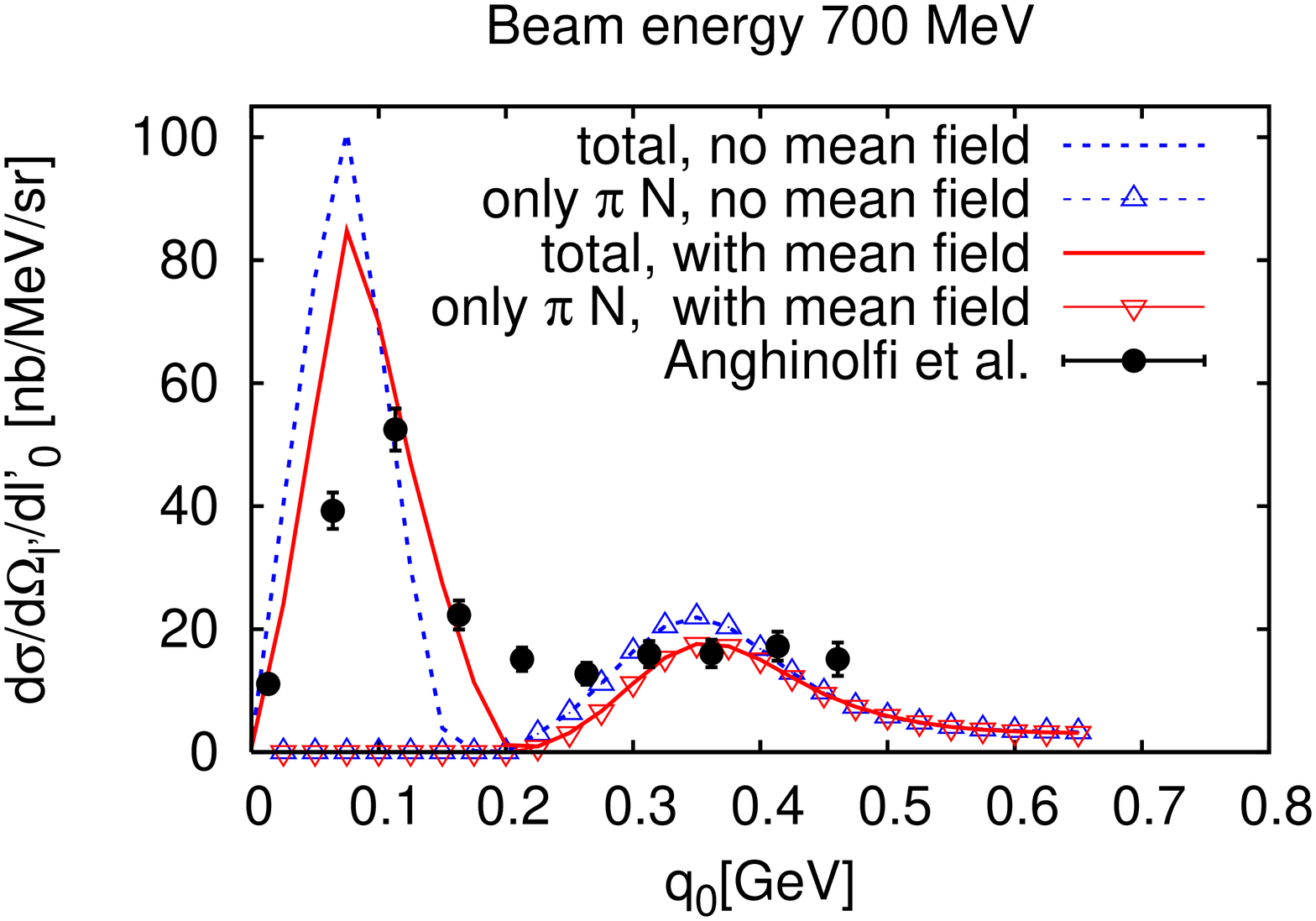}
\includegraphics[width=0.3\textwidth,angle=270]{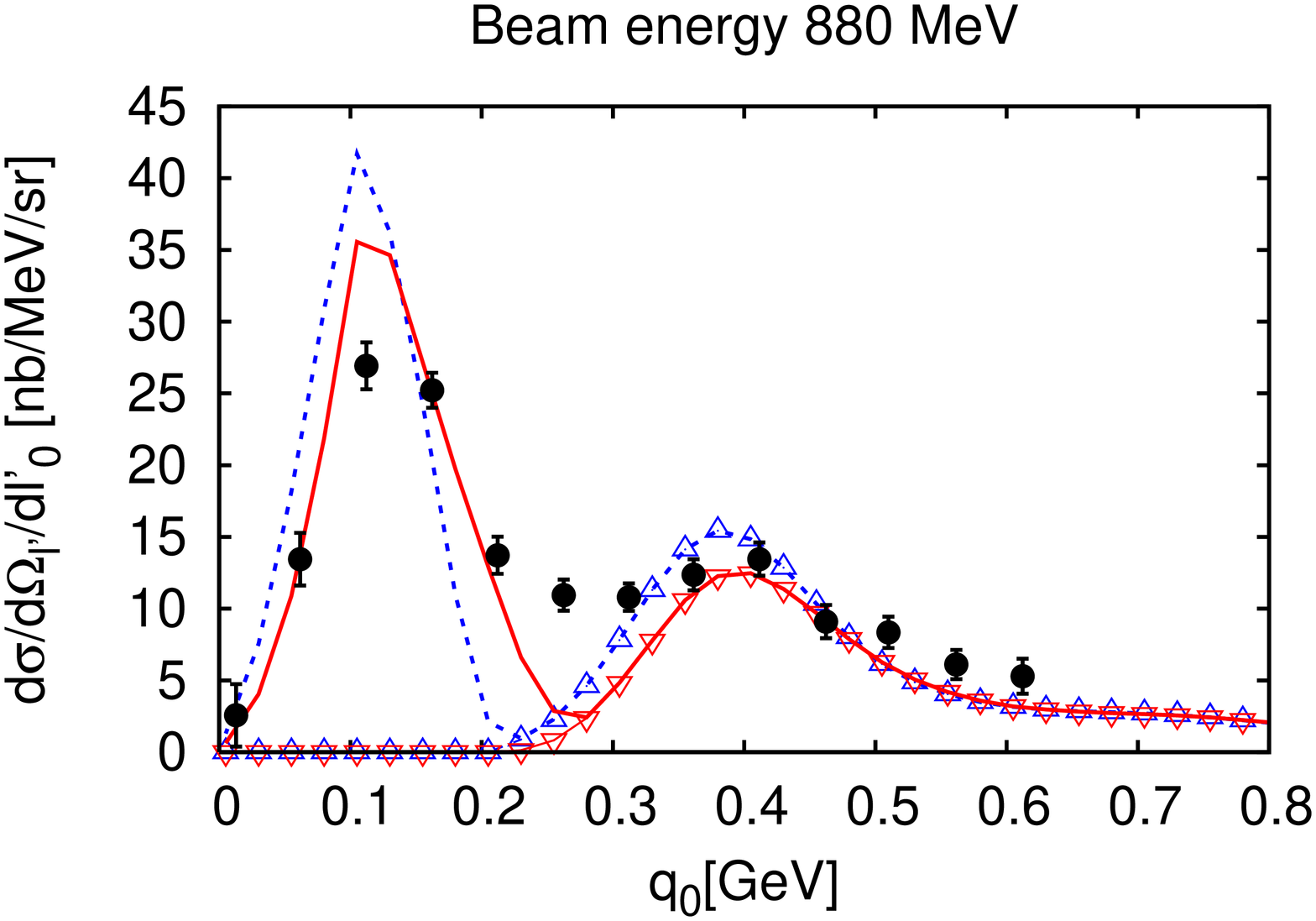} 
\caption{Electron scattering off $^{16}$O at incoming electron energies of 700 MeV and 880 MeV and an electron scattering angle of $37^\circ$. The plot shows a comparison of our results to the experimental data obtained by Anghinolfi et al.~\cite{Anghinolfi:1996vm} in lab coordinates as a function of the virtual photon energy $q^0$. The dashed curves denote the result without a mean field potential for the nucleon and the solid ones with such a potential. The $\gamma^\star N\to \pi N$ contribution to each result is shown with additional symbols.}
\label{elec_results}
\end{figure}
\begin{figure}[h!]
\includegraphics[width=0.3\textwidth,angle=270]{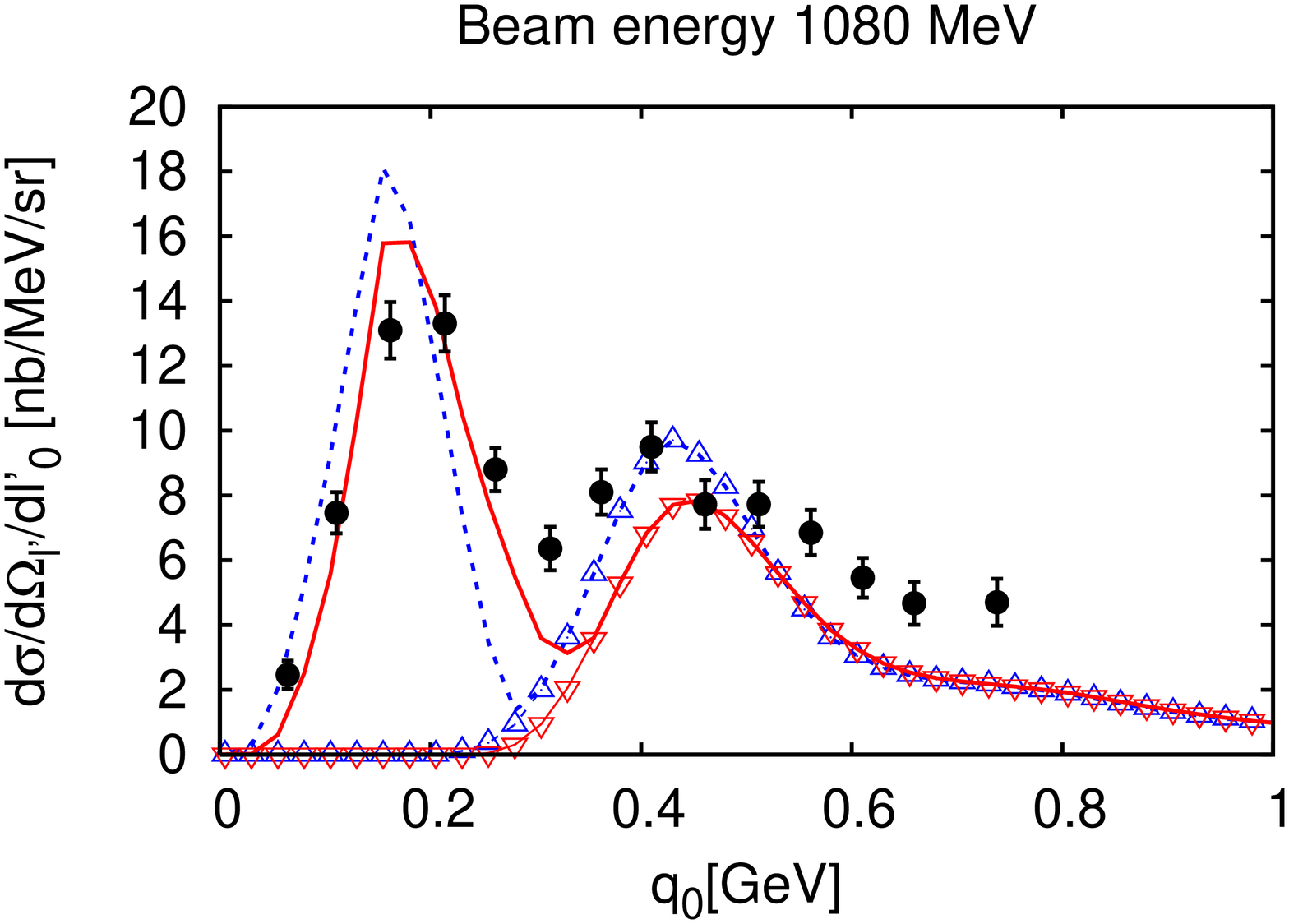}
\includegraphics[width=0.3\textwidth,angle=270]{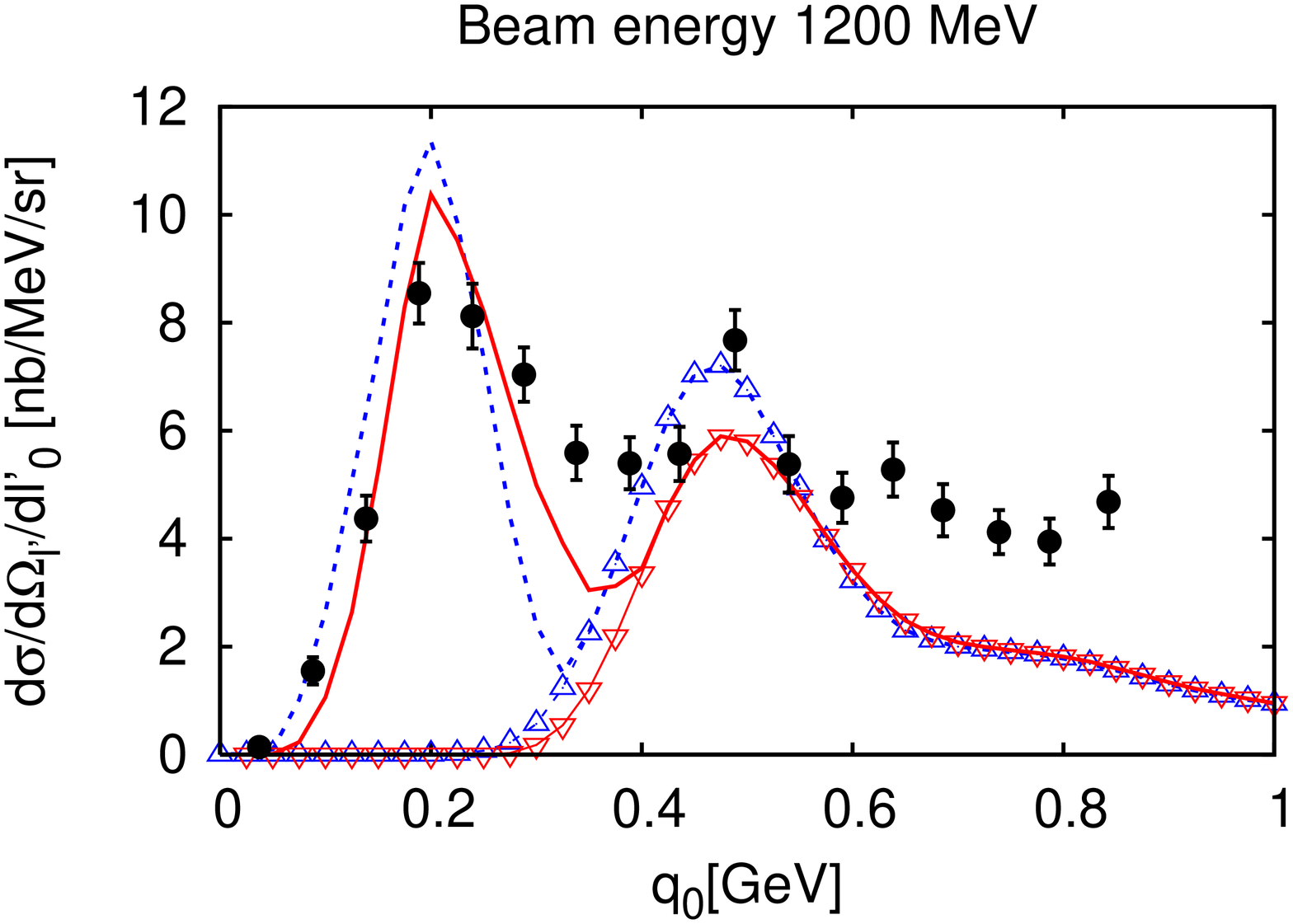}
\includegraphics[width=0.3\textwidth,angle=270]{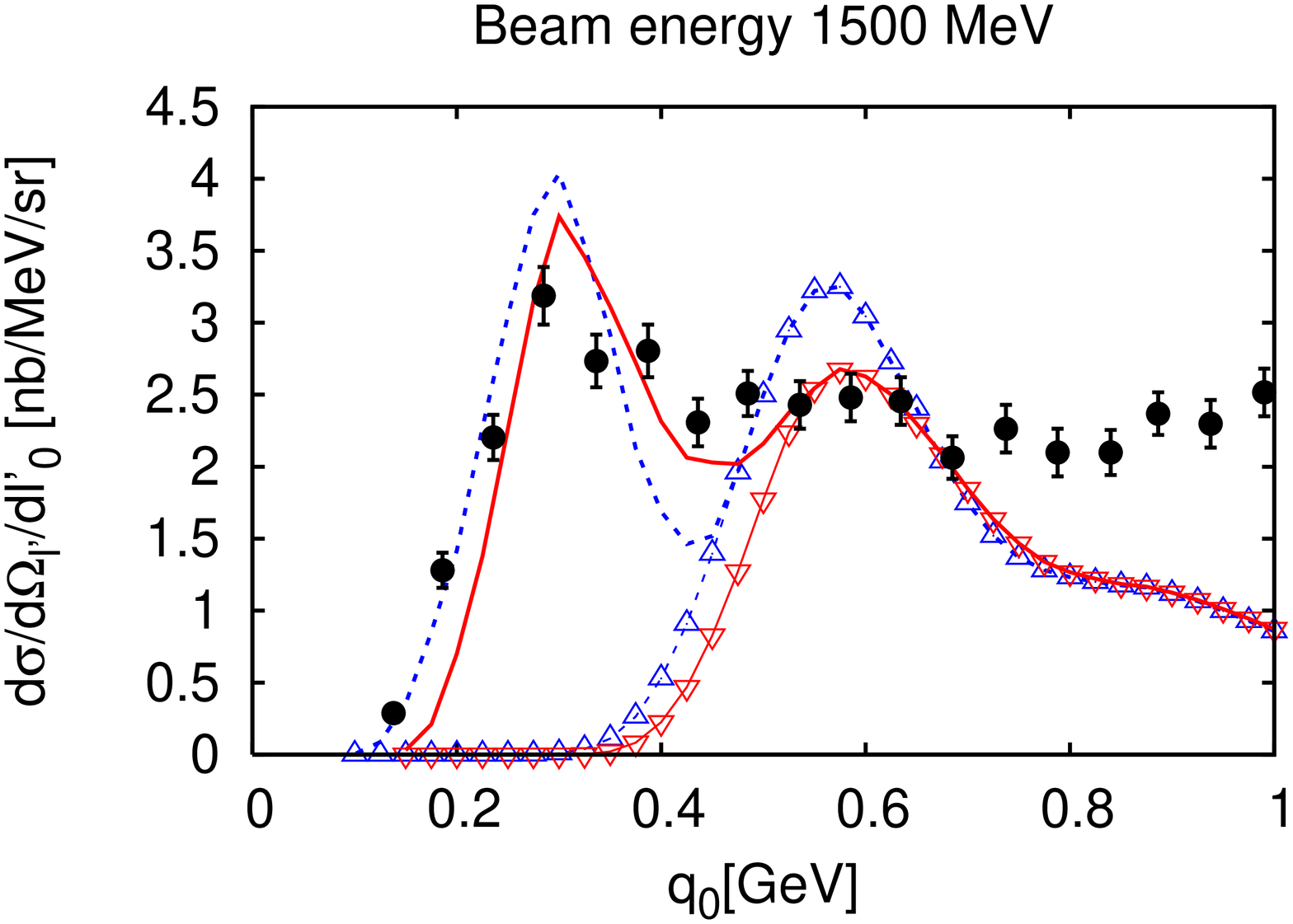}
\caption{Same as fig. \ref{elec_results} at an incoming electron energy of 1080, 1200 and 1500 MeV.}
\label{elec_results2}
\end{figure}
In fig. \ref{elec_results} and \ref{elec_results2} we show a comparison of our inclusive cross section to the experimental data obtained by Anghinolfi et al.~\cite{Anghinolfi:1996vm}. At different beam energies ranging from 700 to 1500 MeV, we show calculations both without (dashed curves) and with (solid curves) a mean field potential for the nucleon. The $\gamma^\star N\to \pi N$ contribution to each result is shown with additional symbols. 

Especially at low energies, there are two distinct peaks in the experimental data: the quasi-elastic peak at low photon energy and the pion-production peak at higher photon energy. At lower energies, we overestimate the experimental data in the quasi-elastic peak and give a good description of the pion production region. Note that especially at low photon energies, nucleon-nucleon correlations and particle-hole excitations (see e.g. \cite{Benhar:2005dj,Gil:1997bm}) and the broadening of the $\Delta$ resonance are expected to play an important role. Up to now, we neglected these effects in our model which might explain this discrepancy. Going to higher energies we observe a better agreement in the quasi-elastic peak and a lack of strength above the one-pion production peak. Here, we are at the present stage simply missing more energetic channels as e.g. $\gamma^\star N \to N \pi \pi$. The overall agreement improves when the nucleon potential is taken into account.

\section{Summary}
We have studied pionic double charge exchange on different nuclear targets ($^{16}O$, $^{40}Ca$ and $^{208}Pb$) in the $\Delta$ region ($\ekin=120,150,180$~MeV) with a semi-classical couple channel transport model (GiBUU). Furthermore, we compared the results of our model with the extensive set of data taken at LAMPF \cite{Wood:1992bi}, achieving a good agreement, not only for the total cross section, but also for angular distributions. The scaling of the total cross sections pointed out in \cite{Gram:1989qh} could be reproduced. However, we found that two important effects that break this scaling: neutron skins and Coulomb forces compensate each other.
We have shown in section \ref{densChapter} that the DCX cross section is very sensitive to the size of the neutron skin. A precise measurement of DCX at forward angles combined with a theoretical analysis could bring quantitative results on the neutron skins of heavy nuclei. 
We conclude that the implementation of pion re-scattering and absorption in the GiBUU transport model successfully passes the demanding test of describing double charge exchange reactions. \newla{Thus the semi-classical approach is well suited to describe pion dynamics in nuclei for pion kinetic energies greater $\ekin\approx 30\MeV$.}

As a second application we have presented our treatment of electron scattering off nuclei. In this problem we have employed in-medium kinematics and parametrized the form factors for QE-scattering according to Budd et al.\cite{Budd:2003wb} and Krutov et al.~\cite{Krutov:2002tp} and used the $\pi$-production form-factors of the MAID\cite{MAIDWebsite,Tiator:2006dq,Drechsel:1992pn} analysis. We have evaluated so far only inclusive cross sections and compared to data by Anghinolfi et al.~\cite{Anghinolfi:1996vm}. In comparison to the data we found good agreement in the pion production energy regime, whereas the quasi-elastic scattering is overestimated for low photon energies. In future we plan to extent the description to exclusive channels, such as e.g. single $\pi$ production, and plan to include more energetic channels like e.g. $\pi\pi$ production and a more realistic spectral function for the nucleon~\cite{Froemel:2003dv,Konrad:2005qm}.

\begin{acknowledgments}
The authors thank Steven A. Wood for his cooperation by promptly  retrieving for us the experimental data for DCX. We also thank Lothar Tiator for making a compilation of the MAID form factors available to us and for his kind support. We are grateful to all members of the GiBUU group, especially Kai Gallmeister and Tina Leitner. This work was supported by Deutsche Forschungsgemeinschaft (DFG). 
\end{acknowledgments}


\bibliography{literatur}

\end{document}